\documentclass[11pt]{article} 
\usepackage{graphicx, epstopdf, fullpage, amsmath, amssymb, subfigure, abstract,capt-of} 
\usepackage[parfill]{parskip} 
\usepackage[T1]{fontenc}

\usepackage[super,comma]{natbib} 
\usepackage{titling} 
\usepackage{setspace}

\usepackage[labelfont={bf}, margin=1cm]{caption}

\newcommand{\figref}[1]{(Fig.~#1)}

\newcommand{\argmax}{\mathop{\scriptstyle \arg\!\max}} 
\newcommand{\ud}{\,\mathrm{d}} 
\newcommand{\comment}[1]{}

\newcommand{\argmin}{\mathop{\scriptstyle \arg\min}}

\makeatletter 
\def\@cite#1#2{[\if@tempswa#2\fi#1]} 
\makeatother

\long
\def\symbolfootnote[#1]#2{
\begingroup

\def\thefootnote{\fnsymbol{footnote}}\footnote[#1]{#2}
\endgroup}

\title{\bf\Large Neural and perceptual signatures of efficient sensory coding \\[4ex]} 
\author{\bf Deep Ganguli and Eero P. Simoncelli\\[2ex]
Howard Hughes Medical Institute \\
Center for Neural Science, and\\
Courant Institute of Mathematical Sciences \\
New York University\\
New York, NY 10003 \\[1.5ex]
\texttt{\{dganguli,eero\}@cns.nyu.edu} \\
} 
\date{}

\begin{document}

\maketitle 

\renewcommand{\abstracttextfont}{\normalfont} 
\renewcommand{\abstractname}{Summary}

\noindent{\bf The mammalian brain is a metabolically expensive device, and evolutionary pressures have presumably driven it to make productive use of its resources. For sensory areas, this concept has been expressed more formally as an optimality principle: the brain maximizes the information that is encoded about relevant sensory variables, given available resources \cite{Attneave1954, Barlow1961,Simoncelli2001}. Here, we develop this efficiency principle for encoding a sensory variable with a heterogeneous population of noisy neurons, each responding to a particular range of values. The accuracy with which the population represents any particular value depends on the number of cells that respond to that value, their selectivity, and their response levels. We derive the optimal solution for these parameters in closed form, as a function of the probability of stimulus values encountered in the environment. This optimal neural population also imposes limitations on the ability of the organism to discriminate different values of the encoded variable. As a result, we predict an explicit relationship between the statistical properties of the environment, the allocation and selectivity of neurons within populations, and perceptual discriminability. We test this relationship for three visual and two auditory attributes, and find that it is remarkably consistent with existing data.}

Neurons in sensory systems are commonly characterized in terms of their selectivity or ``tuning'' for particular stimulus variables (e.g., acoustic frequency, or visual orientation). It is generally believed that for any given variable, populations of cells, each with different selectivity, and arranged so as to cover the full range of stimulus values, form a distributed representation or code for that variable within the brain.\comment{***EPS: feels like we should cite something, but not sure what. }  These codes are believed to both enable, and limit, subsequent perception.\comment{**EPS: again}  In particular, the ability of an observer to discriminate stimuli that differ in terms of these same stimulus variables depends on the number, selectivity, and noise properties of neurons in the underlying neural population \cite{Shadlen1996}.
\comment{
@article{Shadlen1996,
author = "M N Shadlen, K H Britten, W T Newsome, J A Movshon",
title = "A Computational Analysis of the Relationship between
Neuronal and Behavioral Responses to Visual Motion",
journal = "jneurosci",
month = "Februrary",
volume = 16,
number = 4,
pages = "1486--1510",
year = 1996}
}

Nearly all theoretical analyses of neural population codes have used homogeneous populations, in which the neurons cover the stimulus space with tuning curves that are identical shifted copies of a common function \cite{Seung1993,Zemel1998,Zhang1999,Pouget1999,Dayan2001}.  For many sensory variables, however, measurements of neural population properties suggest that the representation is heterogeneous.  \comment{***EPS: GIVE EXAMPLES}
Moreover, measured perceptual discrimination thresholds for many sensory variables also exhibit heterogeneity.
A common example arises in the form of Weber's law, in which discriminability is found to be proportional to stimulus value.  
\comment{***EPS: GIVE more EXAMPLES.}  
Despite the ubiquity of neural and perceptual heterogeneity, no current theory explains why this should be the case, or how these heterogeneities might be related to each other.

We have recently proposed that these variations could arise when the tuning curves of a neural population are arranged so as to maximize the information transmitted about stimuli that are heterogeneous in their frequency of occurrence \cite{Ganguli10c,Ganguli12b}.  Here, we show that this theory provides remarkably accurate predictions of neural and perceptual heterogeneity for a variety of different sensory attributes.
Consider a stimulus variable, $s$, that is to be encoded in the responses of a population of $N$ noisy neurons.  For simplicity, assume that the neuronal response variability is Poisson-distributed with a rate parameter defined by the tuning curve\cite{Seung1993}, and that for any stimulus, the responses of the neurons in the population are uncorrelated. \footnote{The derivation can be generalized to a class of distributions that include correlations\cite{Ma2006} without changing the form of the result \cite{Ganguli12b}.}  In addition to fixing the number of neurons, we also restrict the 
the total spike rate of the population to a maximum value of $R$, reflecting a limitation on metabolic resources \cite{Lennie03,Laughlin98}.

The presence of neural response noise, in concert with the two resource constraints, place limits on the accuracy with which stimuli can be represented by the population.   The literature on population coding has thoroughly examined this issue in the case of a homogeneous population.  If one assumes that the population covers the stimulus space, with adjacent cells overlapping so as to leave no gaps, the accuracy with which a decoder can recover the stimulus value from the population response \footnote{Technically, the result is stated in terms of the Fisher information, which provides a bound on the variance of any unbiased estimator that attempts to recover the stimulus value from the population \cite{Cox1974}.  The Fisher  information also provides a bound on the perceptual discriminability that can be achieved using any estimator, even one that is biased \cite{Series2009}.}
scales inversely with $N^2 R$.  Thus,
the accuracy of the representation improves when either $N$ or $R$ are increased. 

\begin{figure}
\begin{center}
{
\includegraphics[scale=1]{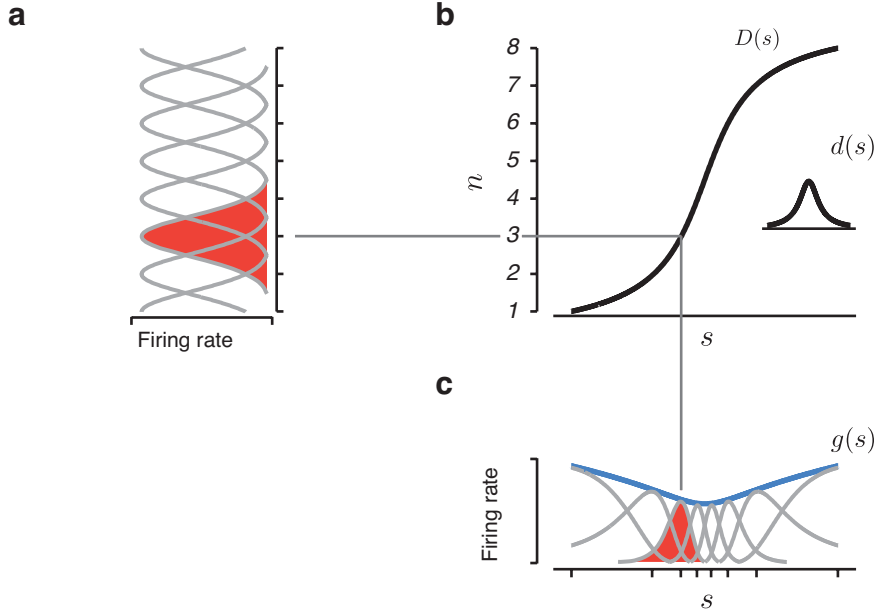}} 
\end{center}
\caption{{\bf Heterogeneous population parameterization.} {\bf a,} An initial homogeneous population containing $N$ cells with shifted copies of a bell-shaped tuning curve, $u(s)$, arranged such that the cells evenly ``tile'' the unit lattice: $u_n(s) = u(s-n)$ such that $\sum_{n=1}^{N}u(s-n) \approx 1$. {\bf b,} The space is non-uniformly warped so as to achieve a desired cell density, $d(s)$, which specifies the number of cells per unit stimulus (shown in inset). The integral of this density function, $D(s) = \int_{-\infty}^{s}d(t)\ud t$, provides a warping function that achieves the desired density, resulting in a set of heterogeneous tuning curves: $h_n(s) = u(D(s)-n)$. The entire population is warped together, ensuring that the width of each tuning curve is inversely proportional to the cell density, and that the warped population will also cover the stimulus space: $\sum_{n=1}^{N}u(D(s)-n) \approx 1$. {\bf c,} The heterogeneous population (with tick marks denoting the preferred stimulus of each cell, $s_n = D^{-1}(n)$) is multiplied by the gain function, $g(s)$ (blue), which controls the maximum average firing rate of each cell. A single tuning curve is highlighted (red) to illustrate the effect of the warping and scaling operations. The total resource constraints ($N$ cells and $R$ total spikes) can be expressed in terms of the cell density function: $N = \int d(s) \ud s$, and gain: $R = \int p(s) g(s) \ud s$.}
\label{parameterization} 
\end{figure}

Suppose the environment is inhomogeneous, in that the frequency of occurrence of a stimulus variable, expressed as a probability distribution, $p(s)$, varies significantly over the range of $s$. Intuitively, a ``good'' sensory system would allocate a higher proportion of neurons or spikes (or both) to the most frequently occurring stimuli, improving the encoding accuracy of those stimuli at the cost of decreasing the encoding accuracy of infrequently occurring stimuli.  Is there a choice that is optimal?  In order to answer this question, we parameterize a nonuniform allocation of neurons using a continuous function, $d(s)$, that represents the {\em cell density}.  The  heterogeneous population is formed by using this density function to warp a homogenous population, as depicted in \figref{1}.    An important feature of this parameterization is that it  enforces an inverse relationship between tuning widths and cell density, 
and thus preserves the relative overlap between adjacent tuning curves.  If the homogeneous population is chosen so that the tuning curves cover  their stimulus space with modest overlap, then the warped heterogeneous population will do the same.
Similarly, we parameterize a nonuniform allocation of spikes using a continuous {\em gain} function, $g(s)$, which is simply multiplied by each of the warped tuning curves.   Under this local parameterization of resource allocation, the accuracy of the representation scales inversely with $d(s)^2 g(s)$, analogous to the homogeneous case \cite{Ganguli10c}.

This parameterization allows us to optimize the population for the transmission of stimulus information.  Specifically, 
the average of the log accuracy provides a lower bound on transmitted information\cite{Brunel1998}, and thus we need to solve the following constrained optimization problem:
\begin{equation*}
\argmax_{d(s),g(s)} \int p(s) \log \left(d^2(s)g(s)\right) \ud s, \qquad\text{subject to} \quad \int d(s) \ud s = N, \quad\text{and} \quad \int p(s)g(s) \ud s = R,
\end{equation*}
A closed form solution is readily obtained using calculus of variations:
\begin{equation}
d(s) = N p(s), \qquad w(s) \propto \frac{1}{d(s)} = \frac{1}{Np(s)}, \qquad g(s) = R. \label{eq:optimum} 
\end{equation}

The structure of the optimally efficient population directly reflects the statistical properties of the environment, as specified by $p(s)$. Specifically, the cell density is proportional to the stimulus distribution, ensuring that frequently occurring stimuli are encoded with greater precision, using a larger number of cells with correspondingly narrower tuning. On the other hand, we see that the maximal response (gain) of the cells in the optimal population is constant, independent of the preferred stimulus value. Since we have assumed the tuning widths are inversely proportional to cell density, and thus to the stimulus distribution, this solution implies that the average response of each neuron (over stimuli encountered in the world), is identical for all neurons in the population. Finally, the unknown total resource values $\{N, R\}$ appear only as multiplicative scale factors in the expressions for gain and density, and thus the optimal solution provides a unique and testable predictions for the shapes of both the cell density and tuning width as a function of preferred stimulus.

Finally, substituting the optimal cell density and gain into the expression for accuracy given above allows us to make a prediction about the minimum discriminination thresholds that could be achieved based on this population :
\begin{equation}
\delta_{\rm min}(s) \propto \frac{1}{\sqrt{d^2(s) g(s)}} = \frac{1}{N \sqrt{R} p(s)} \label{eq:discriminability} 
\end{equation}
The solution predicts that frequently occurring stimuli should be more discriminable (specifically, inverse discrimination thresholds should be proportional to the probability of encountering a stimulus value). The shape of this solution is again a simple function of the stimulus probability, $p(s)$, scaled by a multiplicative factor that depends on neural resources and an additional factor that depends on the experimental conditions under which discrimination thresholds are measured (e.g., criterion value, stimulus duration, or intensity). As a result, the solution provides a unique prediction of the shape of perceptual discrimination as a function of stimulus value.

\begin{figure}
\begin{center}
{
\includegraphics[scale=.91]{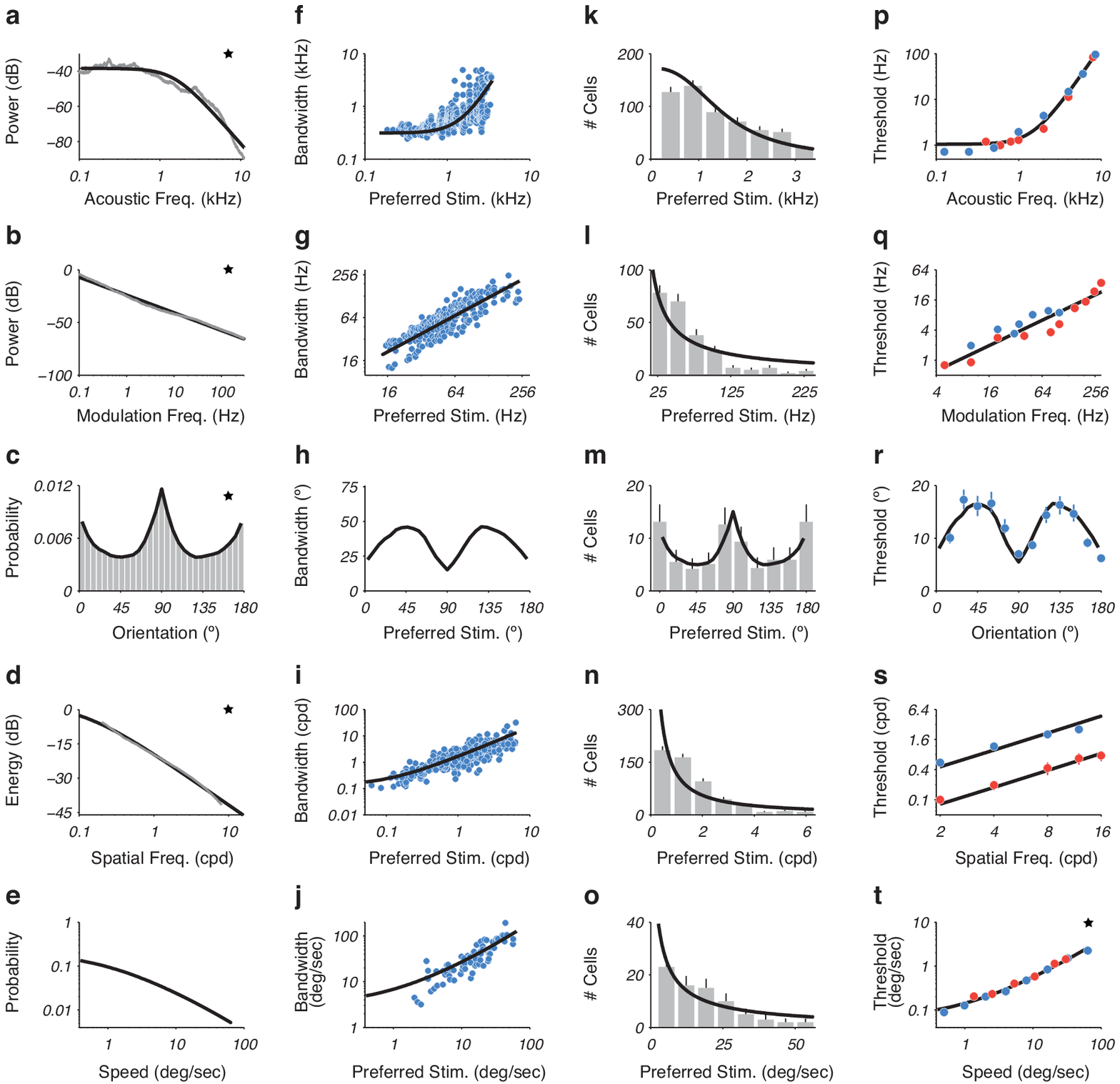}} \label{predictions} 
\end{center}
\end{figure}
\begin{figure*}
\caption{{\bf Tests of theoretical predictions.} Comparison of predicted relationship between environmental distributions, neural population properties, and psychophysical discrimination thresholds to data. Each row corresponds to a particular sensory attribute: acoustic frequency, modulation frequency, local orientation, spatial frequency, and speed. {\bf a-e,} Environmental distributions. {\bf f-j,} Tuning widths as a function of preferred stimulus, for neural populations known to be tuned for each attribute. {\bf k-o,} Histograms of the preferred stimulus values for all neurons in each population. {\bf p-t,} Discrimination thresholds, averaged across multiple human subjects. 
For each row, the data in the starred panel was fit with a parametric form or histogram density estimate (thick black lines, see Methods for fitting details). These curves were transformed according to Eqs. (\ref{eq:optimum}) or (\ref{eq:discriminability}) to generate predictions for all other panels in the same row. Since the predictions include an unknown scale factor (which is a function of the resources, $N$ and $R$), each curve is rescaled to minimize the squared error of the associated data. Environmental distributions were estimated from ensembles of images or sounds (see Methods for details): {\bf a-b,} Distributions of acoustic and modulation frequencies computed from a compilation of animal vocalizations, background sounds\cite{Storm1994, Storm1994a,Emmons1997}, and recordings made while walking around a suburban university campus\comment{McDermott????}.
 {\bf c-d,} Distributions of local orientations and spatial frequencies, respectively, computed from a compilation of three large databases of images containing both natural and man-made scenes\cite{Hateren1998, Doi2003,Olmos2004}. Physiological data (tuning widths and cell densities) were estimated from published data sets: {\bf f,k,} $553$ auditory nerve fibers measured in cats\cite{Carney1999}, adapted from\cite{Smith2006}. {\bf g,l,} $262$ neurons in the central nucleus of the inferior colliculus measured in cats\cite{Rodriguez2010}. {\bf h,m,} $79$ orientation-tuned V1 simple cells recorded foveally\cite{Mansfield1974} (note: data set does not include tuning widths, so only the theoretical prediction is shown). {\bf i,n,} $538$ neurons in Macaque primary visual cortex (V1)\cite{Cavanaugh2002}. {\bf j,o,} $76$ speed-tuned cells in Macaque middle temporal cortex (MT)\cite{Movshon15}. Perceptual discrimination thresholds were taken from published data sets: {\bf p,} Acoustic frequency data from two different studies shown in red\cite{Moore1973} and blue\cite{Wier1977}. {\bf q,} Modulation frequency data from two different studies shown in red\cite{Formby1985} and blue\cite{Lemanska2002}. {\bf r,} Orientation discrimination data \cite{Girshick2011}. {\bf s,} Spatial frequency discrimination thresholds measured with sinusoidal gratings at $10\%$ contrast (red)\cite{Caeli1983}, and $25\%$ contrast (blue)\cite{Regan1982} (note that theoretical predictions are scaled differently, to allow for differences in stimulus conditions). {\bf t,} Speed discrimination data from two different studies shown in red\cite{McKee1984} and blue\cite{DeBruyn1988}.}
\end{figure*}

Efficient population coding predicts explicit relationships between sensory statistics, physiological tuning properties, and perceptual discriminability (Eqs.~\ref{eq:optimum}\,\&\,\ref{eq:discriminability}). We tested these relationships in the context of two auditory attributes (acoustic frequency and modulation frequency), and three visual attributes (local orientation, spatial frequency, and retinal speed) \figref{2}. Each of these attributes exhibits substantial heterogeneity in their statistical, physiological, and perceptual representations. Data in the first column \figref{2a-e} correspond to stimulus distributions for each attribute, as estimated from large databases of photographic images or sounds obtained from natural environments (see Methods). Physiological data \figref{2f-j} are taken from single-cell electrophysiological recordings in primate or cat that report the independently measured tuning widths of a large population of neurons as a function of their preferred stimuli. These measurements were gathered across multiple animals, and in  a diverse set of brain areas (the auditory nerve fibers, the inferior colliculus, the primary visual cortex, and the middle temporal cortex), each chosen based on a substantial literature identifying the tuning properties of those neurons for the stimulus feature of interest. For the case of local orientation, we also analyzed another physiological data set in which tuning widths are reported (see  Appendix). Estimates of the cell density in each area \figref{2k-o} are obtained with a histogram binned over the preferred stimuli. Discrimination thresholds for each sensory attribute \figref{2p-t} were measured in human perceptual experiments. In some cases, data are reported from two perceptual data sets (distinguished by color) obtained under different experimental conditions.

For each attribute, we find that the predicted relationships between the environment, physiology, and perception (Fig.~2, thick black lines) are consistent with the data. In most cases, we used the histogram of the environmental data as an estimate of $p(s)$, and then used this to predict the physiological and perceptual data. Predicted curves for tuning width and perceptual discriminability are individually re-scaled to best match the corresponding data (since the true scale factors depend on the unknown values of $N$ and $R$). In the case of local image speed \figref{2e,j,o,t} for which environmental data are technically difficult to estimate, we used the theory in reverse, fitting the perceptual data and using this to generate a prediction for $p(s)$. For all other sensory attributes, we find that predictions of perceptual discrimination data are remarkably accurate. For all attributes, both physiological predictions are well supported by the data and confirm our assumption that tuning widths are inversely proportional to cell density. For tuning widths, the predictions account for $(39.6,69.1,47.7,67.4)\%$ of the variance in the data, which corresponds to $(95.8, 93.5, 97.1,98.8)\%$ of the data variance that is accounted for by the best-fitting power law with two additional free parameters. Estimates of cell density may not be appropriately represented in the data, as they are limited by sample size, uncontrolled variables (such as eccentricity for visual neurons) and potential biases in electrode sampling (see Appendix). Finally, we examined the gain of several neural populations (see Appendix). We find that, although there is significant variation in these values across the population, it exhibits no systematic relationship to the stimulus value.

\begin{figure}
\begin{center}
{
\includegraphics[scale=1.2]{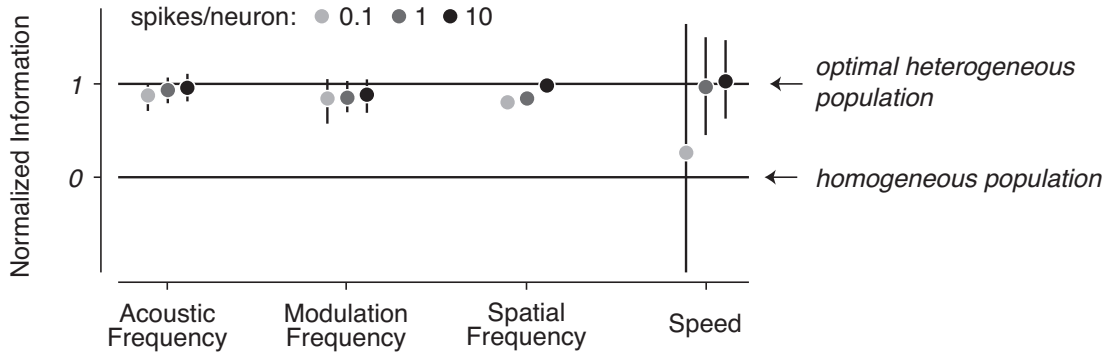}} 
\end{center}
\caption{{\bf Information transmitted by the physiologically measured neural populations.} Normalized information for data sets in Fig.~2f,g,i,j computed for three different average firing rates, relative to a homogeneous population and an optimal heterogeneous population (all matched for number of cells and average firing rate). Error bars denote the $5$th and $95$th percentiles estimated from $1000$ bootstrap re-samplings of the data.}
\label{information} 
\end{figure}

The neural data exhibit scatter about their predicted values, and we wondered how much this variability might degrade the transmitted information. To quantify this, we compared the amount of information transmitted by each observed neural population, to that transmitted by the theoretically optimal populations (see Methods). We find that, relative to a homogeneous population with the same resources, the observed neural populations encode most of the information that would be transmitted by the optimal population \figref{3}. Thus, despite the variability of the physiological data, we conclude that neural populations are near-optimal in their efficiency for transmitting signals drawn from their associated environmental distributions.

The notion that an organism is adapted both physiologically and perceptually to the statistics of the natural environment is of fundamental importance to evolutionary biology. We have developed a precise instantiation of this notion, and used it to derive a novel set of predictions relating environmental distributions, the tuning of properties neural populations, and perceptual discriminability.  We showed that, despite the simplicity of the formulation, the resulting predictions are supported by physiological data from a diverse set of brain areas, as well as human perceptual data.

Perhaps the strongest assumption our formulation is the one embedded in the parameterization: the tuning curves are obtained by warping a homogeneous population to achieve a desired cell density.  It is this assumption that allows for the closed form solution to the optimization problem - without it, the solution is under constrained.
The assumption is based on the intuitively sensible notion that the optimal shape and overlap of tuning curves, whatever it is, should be the same regardless of cell density.  
Comparison of the measured cell density and tuning widths shows that this assumption is consistent with the physiological data, although one would like to see confirmation on a single population, measured in a single animal.
Finally, it is worth noting that if one takes the warping solution seriously as a physiological model (as opposed to merely a useful computational trick), it makes additional predictions regarding the skewed shapes of tuning curves (e.g., see Fig. 
\ref{parameterization}).   Details of tuning curve shapes are not usually reported in physiological literature, but qualitatively appropriate asymmetries in tuning curves have been noted in some cases \cite[e.g.,~]{Nover2005}.

Our results generalize and extend a number of previously published results on efficient sensory coding. 
For populations of identical tuning curves, a non-uniform distribution of preferred stimuli was found to be optimal\cite{Brunel1998, Harper2004}. However, these results were not compared directly to physiological data, and because they assumed uniform tuning widths, cannot account for perceptual discriminability \figref{2p-t}. A number of studies have examined coding efficiency for single neurons with {\em monotonic} response functions, demonstrating that the optimal solution is proportional to the cumulative stimulus distribution\cite{Laughlin1981,Nadal1994,McDonnell2008}, consistent with some physiological data\cite{Laughlin1981}. 

As with nearly all previous studies of neural population coding, our results are limited to the description of single stimulus variables, and generalization to the joint encoding of multiple attributes is not straightforward. Nevertheless, we find that our physiological predictions are consistent with populations of linear receptive fields are that are numerically optimized to encode ensembles of natural images\cite{Olshausen1996,Bell1997} or sounds\cite{Smith2006}. Specifically, the tuning characteristics of these optimized receptive fields are consistent with our predictions of cell density and tuning width for the encoding of orientation, spatial frequency, and acoustic frequency (see Appendix).

Our predictions arise from the principle of coding efficiency, 
which many have argued is a reasonable objective for early stages of sensory processing, but seems unlikely to explain more specialized later stages such as those responsible for producing actions\cite{Geisler2009}.   
But for the cases tested here, when we examined the predictions of an alternative theoretical proposal $-$ that neural systems are optimized for discrimination performance\cite{Twer2001} - we found that they were far less consistent with the data (see Appendix).  Thus, our results provide additional support to the notion that information provides the strongest generic criterion for sensory system design.   

An additional unexpected, but powerful benefit of our optimal population solution emerges in the context of decoding, or ``readout'' of the population. If we take seriously the notion that perception is a process of inference \cite{Helmholtz2000}, then later stages of processing must be aware not just of the noise properties of the representation \cite{Ma2006,Jazayeri2006}, but must also have knowledge of the frequency of occurrence of sensory attributes in the environment. Although such prior information has been widely used in formulating Bayesian explanations for perceptual phenomena \cite{Knill1996}, the means by which it is represented within the brain is currently unknown \cite{Simoncelli2009}. Our results provide a potential solution, in which prior probabilities are implicitly embedded in the arrangement and selectivity of tuning curves. The responses of an efficient population can be decoded in a biologically plausible mechanism that correctly combines current sensory information with the embedded prior \cite{Ganguli12a}. Thus, an efficient coding strategy may offer unforeseen benefits for explaining later stages of sensory processing.

\comment{Finally, although our analysis has focused on the encoding of sensory information in neural populations, the results have direct implications for the decoding of the population responses. In particular, if one takes seriously the notion that perception is a process of inference\cite{Helmholtz2000}, then subsequent decoding stages must incorporate knowledge of the uncertainty of incoming sensory information, as well as the frequency of occurrence of sensory attributes in the environment. Recent publications has explored the representation of sensory uncertainties \cite{Jazayeri2006,Ma2006}, while others have proposed that environmental probabilities could be represented through use of heterogeneous cell populations\cite{Simoncelli2009,Fischer2011,Girshick2011}. Our results provide a bottom-up principle for a specific form of such an encoding, in which prior probabilities are implicitly embedded in the arrangement and selectivity of tuning curves. The responses of such a population can be decoded using a biologically plausible mechanism that correctly combines noisy sensory information with the embedded prior\cite{Ganguli12a}. Thus, efficient population representations may offer unforeseen benefits for explaining subsequent stages of sensory processing.}

\section*{Methods}

\subsection*{Fisher information and resource constraints for a heterogeneous population} 
The Fisher information, $I_u(s)$, for a uniform population \figref{1a} with independent Poisson responses is:
\begin{equation*}
I_u(s) = \sum_{n=1}^{N} \frac{u'^2(s-n)}{u(s-n)} = \sum_{n=1}^{N} \phi(s-n) = I_u, 
\end{equation*}

where $u'(s-n)$ is the derivative of the $n^\text{th}$ tuning curve\cite{Seung1993}. We assume that the uniform tuning curves, and their Fisher kernels, $\phi(s-n)$, evenly ``tile'' the stimulus space: $\sum_{n=1}^{N}u(s-n) = 1$, and $\sum_{n=1}^{N} \phi(s-n)= I_u$. The heterogeneous population is formed by warping the tuning curves according to the cell density function \figref{1b}, which preserves these tiling properties: $\sum_{n=1}^{N}u(D(s)-n) = 1$, and $\sum_{n=1}^{N} \phi(D(s)-n)= I_u$. Each tuning curve, $h_n(s)$, is then scaled by a gain factor, $g_n$, that controls the maximum average firing rate of the cell: $h_n(s) = g_nu\left(D(s)-n\right)$. We define a smooth continuous version of the gain as: $g(s) = \sum_{n=1}^{N} g_n u\left(D(s)-n\right)$ \figref{1c} taking advantage of the fact that the tuning curves tile. The Fisher information for the heterogeneous population may then be expressed in terms of the cell density and gain\cite{Ganguli10c} as:
\begin{equation}
I_h(s) = d^2(s)\sum_{n=1}^{N} g_n\phi(D(s)-n) \approx d^2(s)g(s)I_u, \label{eq:heterofish} 
\end{equation}

where we have assumed that $\sum_{n=1}^{N} g_n \phi\left(D(s)-n\right) \approx g(s)I_u$. For this parameterization, the constraint on the total number of spikes is: $R = \int p(s) \sum_{n=1}^{N}g_nu(D(s)-n) \ud s = \int p(s) g(s) \ud s$, and the constraint on the number of neurons is $N = D(\infty)$, or $N = \int d(s) \ud s$.

\subsection*{Fisher bound on mutual information} 
The mutual information between the stimulus and the population response is bounded from below by the Fisher information\cite{Brunel1998}:
\begin{align}
I(\vec{r};s) &= H(s) + \int \int p(\vec{r},s) \log \left(\frac{p(\vec{r}|s)p(s)}{p(\vec{r})}\right) \ud s \ud \vec{r} \label{eq:mutualinfo} \\
& \geq H(s) + \frac{1}{2}\int p(s) \log \left(\frac{I_ud^2(s)g(s)}{2\pi e}\right) \ud s. \label{eq:fisherapprox} 
\end{align}

We verified that this bound is tight by numerically computing and comparing the two quantities for the physiological and environmental conditions reported here. We used estimates of the stimulus distributions, $p(s)$, and cell densities for each attribute (Fig $2$a-o thick black lines). We assumed an independent Poisson noise model for $p(\vec{r}|s)$, witha $N$ set to the sample size of each population, and varied the total number of spikes per neuron in a metabolically relevant regime: $0.1, 1, \text{ and } 10$ spikes per neuron\cite{Lennie2003}. The stimulus entropy, $H(s)=-\int p(s)\log p(s) \ud s$, and the integral in the Fisher approximation (Eq. \ref{eq:fisherapprox}) were computed by numerical integration. The multidimensional integral in Eq. (\ref{eq:mutualinfo}) was computed via Monte-Carlo integration with $10,000$ samples from the joint distribution $p(\vec{r},s)$\cite{Yarrow2012}. We found that the bound was always within $1\%$ of the Monte-Carlo estimate of the Shannon information. We conclude that our efficient population solution, which optimizes the Fisher bound, is near-optimal for the Shannon information.

\subsection*{Estimating Environmental Distributions}
The environmental distributions for the two auditory attributes (acoustic frequency and modulation frequency) were computed from commercially available compilations of animal vocalizations ($58$ min)\cite{Storm1994,Storm1994a}, background environmental sounds ($113$ min)\cite{Emmons1997}, and recordings made while walking around a suburban university campus ($62$ min). The campus sounds were recorded with a Sennheiser omnidirectional microphone (ME62) and a Marantz solid state recorder (PMD670). The distributions for local orientation and spatial frequency were computed from three publicly available image databases comprised of a total of $816$ natural scenes\cite{Hateren1998,Doi2003,Olmos2004}. \comment{The distribution for speed was predicted according to the theory from the psychophysical data.}

\subsubsection*{Acoustic Frequency} 
We used the ensemble power spectral density as an estimate of the probability of acoustic frequencies occurring in the natural environment. We computed the power spectral density for each sound file in the database using Welch's method\cite{Welch1967}, with non-overlapping $500$ millisecond segments windowed with a hamming filter to mitigate boundary effects. The ensemble power spectrum, $S(f)$, was fit to all recordings with a modified power-law\cite{Attias1997}:

\begin{equation}
S(f) = \frac{A}{f_0^p+f^p}. \label{eq:modfiedpowerlaw} 
\end{equation}

The parameters were chosen to minimize squared error to the data ($A = 2.4 \times 10^{6}, f_0 = 1.52 \times 10^{3}, p = 2.61$).

\subsubsection*{Modulation Frequency} 
We used the ensemble modulation power spectral density to estimate the probability of modulation frequencies occurring in the natural environment. Each sound in the auditory database was decomposed into subbands using a physiologically motived bank of $30$ raised cosine filters\cite{McDermott2011}. The center frequencies of the filters were equally spaced on an equivalent rectangular bandwidth ($\text{ERB}_{\text{N}}$) scale, and the filter bandwidths (as a function of center frequency) were comparable to those of the human ear\cite{Glasberg1990}. The temporal envelope of the output of each frequency channel was extracted by computing the magnitude of the analytic signal. The temporal modulation power spectrum was computed by averaging the power spectral density of each envelope across all frequency channels. The modulation spectrum of the envelope of a bandpass filter output is inevitably low-pass (with a cutoff determined by the filter bandwidth). To avoid biasing our measurements of modulation statistics, we only included frequencies below this filter cutoff in our average. The ensemble modulation spectrum was fit to all recordings with a modified power law, (Eq. \ref{eq:modfiedpowerlaw}), with parameters chosen to minimize squared error to the data ($A = 0.06, f_0 = 0, p = 0.84$).

\subsubsection*{Local Orientation} 
We used a Gaussian pyramid\cite{Burt1983} to decompose each image in the database into a spatial scale \comment{ ($2-5$ cycles/deg)} that matched that of the grating stimuli used in the orientation discrimination experiment ($4$ cycles/deg) \figref{2r}\cite{Girshick2011}. The horizontal and vertical gradients, centered on each pixel in the resulting image, were computed with local rotation-invariant $5$-tap derivative filters\cite{Farid2004}. We computed an orientation tensor\cite{Granlund1995}, defined as covariance matrix of the gradients pooled across a local region. A pooling size of $1$ degree was chosen to best match the physiological data, which was measured foveally (between $0-3$ deg retinal eccentricity) \figref{2m,h}\cite{Mansfield1974}. We computed three quantities from the eigenvector decomposition of the orientation tensor: the energy (sum of the eigenvalues), orientedness (ratio of the eigenvalue difference to the eigenvalue sum), and the dominant orientation (angle of the eigenvector with the larger eigenvalue). We formed a histogram of the dominant orientations of all tensors for which the energy exceeded the $68$th percentile of all energies in the database, and the orientedness exceeded $0.8$. \comment{the histogram was converted to a probability distribution by normalizing by the total number of tensors that exceeded the two thresholds. } We verified that the resulting distribution did not change significantly for modest changes in both thresholds.

\subsubsection*{Spatial Frequency} 
We used the radially integrated power spectral density of natural images to estimate the probability of spatial frequencies occurring in the natural environment. The power spectral density for each image in the database was computed by taking the magnitude of the windowed Fourier Transform of the image, and integrating the result over orientation. The units of the power spectrum were converted from cycles per pixel to cycles per degree of visual angle by using the appropriate camera settings for which each image was captured. The ensemble spectra were fit with a modified power law (Eq. \ref{eq:modfiedpowerlaw}) with parameters chosen to minimize squared error to the data ($A = 0.21, f_0 = 0.11, p = 1.14$). The form of the estimated spectrum is approximately $1/f$, consistent with previous studies \cite{Field1987,Ruderman1994}.

\subsubsection*{Speed} 
The psychophysical data were fit with a power law, 
\begin{equation*}
\delta(s) = as^p+b, 
\end{equation*}
with parameters chosen to minimize squared error to the data ($a = 0.05, p = 0.93, b = 0.11$). According to the theory (Eq. \ref{eq:discriminability}), the predicted environmental distribution for speed is estimated as $p(s) \propto \delta(s)^{-1}$ \figref{2e}. This ``slow speed prior'' is qualitatively consistent with previous estimates of the distribution of retinal speed\cite{Dong1995,Roth2007,Stocker2006a}.

\subsection*{Estimating normalized information} 
The normalized mutual information, $I_\text{norm}(\vec{r};s)$, is defined as:

\begin{align*}
I_\text{norm}(\vec{r};s) = \frac{I_\text{dat}(\vec{r};s)-I_\text{hom}(\vec{r}|s)}{I_\text{het}(\vec{r};s)-I_\text{hom}(\vec{r};s)} . 
\end{align*}

Here, $I_\text{hom}(\vec{r};s)$ is the information transmitted by the homogeneous population, $I_\text{het}(\vec{r};s)$ is the information transmitted by the optimal heterogeneous population, and $I_\text{dat}(\vec{r}|s)$ is the information transmitted by the observed neural populations. A value of $0$ indicates the measured neural population transmits as much information as a comparable homogeneous population, and a value of $1$ indicates that the measured neural population transmits as much information as the optimal heterogeneous population. Note that the values can exceed the range $[0,1]$.

To compute the normalized information for each attribute \figref{3}, we first constructed a homogeneous population of equal-width Gaussian tuning curves, evenly spaced across the domain of the sensory prior. The number of tuning curves, $N$, was matched to the number of cells observed for each attribute. We used three different values for the average number of spikes, $\frac{R}{N} \in \{0.1, 1,10\}$ spikes per neuron. The width of the Gaussian was chosen such that, after warping the homogeneous population by the optimal density function, the tuning widths of the resulting optimal heterogeneous population (measured as the full width at half maximum value) were well-matched to the predicted tuning widths (Fig.~2f-j, thick black lines). For the physiological datasets, we did not have access to the empirically measured acoustic or modulation frequency tuning curves. We chose to model them with Gamma functions, with rate and shape parameters chosen to minimize the squared error with the values observed in the data \figref{2f-g}. For spatial frequency and speed, we used the measured tuning curves, which were fit to the data with a log Gaussian function \cite{Nover2005}. These tuning curves exhibited substantial variability in their gains \figref{A2}. To ensure a comparison of information transmission for identical resources, we scaled the gains of tuning curves in each of these populations by a single value such that $R$ was the same as in the corresponding homogeneous and optimal heterogeneous populations.

Given the tuning curves, we estimated the normalized mutual information with Monte Carlo integration with $L = 10,000$ samples. We verified that this was a sufficient number of samples by repeating the procedure with $100,000$ samples and obtaining similar results. We also verified that the precise shape of the tuning curve used to model the physiological data for acoustic and modulation frequency did not significantly influence the normalized information by comparing the analysis with Gaussian and raised cosine tuning functions. $95\%$ confidence intervals for the normalized information were obtained with $1,000$ bootstrap estimates of $I_\text{dat}(\vec{r}|s)$. For all attributes and resource constraints (except for speed, under condition $\frac{R}{N} = 0.1$), the normalized information was significantly greater than $0$ and quite close to $1$, indicating that the observed neural populations are nearly optimal for information transmission despite significant heterogeneity in cell density, tuning width, and (for spatial frequency and speed) gain.

\subsection*{Acknowledgements} We thank Tony Movshon and Helena Wang for providing physiological data, Josh McDermott for providing a database of natural sounds and helpful discussions about auditory coding, and Jeremy Freeman for helpful comments on the manuscript.
\comment{***Other funding? NSF? DG I just have the NYU graduate fellowship}
This work was funded by an HHMI investigatorship to EPS.

\bibliography{zotero,simoncelli,others} 

\begin{thebibliography}{10}
\expandafter\ifx\csname url\endcsname\relax
  \def\url#1{\texttt{#1}}\fi
\expandafter\ifx\csname urlprefix\endcsname\relax\def\urlprefix{URL }\fi
\providecommand{\bibinfo}[2]{#2}
\providecommand{\eprint}[2][]{\url{#2}}

\bibitem{Attneave1954}
\bibinfo{author}{Attneave, F.}
\newblock \bibinfo{title}{Some informational aspects of visual perception}.
\newblock \emph{\bibinfo{journal}{Psychological Review}}
  \textbf{\bibinfo{volume}{61}}, \bibinfo{pages}{183--193}
  (\bibinfo{year}{1954}).

\bibitem{Barlow1961}
\bibinfo{author}{Barlow, H.}
\newblock \bibinfo{title}{Possible principles underlying the transformation of
  sensory messages}.
\newblock \emph{\bibinfo{journal}{Sensory Communication}}
  \bibinfo{pages}{217--234} (\bibinfo{year}{1961}).

\bibitem{Simoncelli2001}
\bibinfo{author}{Simoncelli, E.} \& \bibinfo{author}{Olshausen, B.}
\newblock \bibinfo{title}{Natural image statistics and neural representation.}
\newblock \emph{\bibinfo{journal}{Annual Review of Neuroscience}}
  \textbf{\bibinfo{volume}{24}}, \bibinfo{pages}{1193--1216}
  (\bibinfo{year}{2001}).

\bibitem{Shadlen1996}
\bibinfo{author}{Shadlen, M.}, \bibinfo{author}{Britten, K.},
  \bibinfo{author}{Newsome, W.} \& \bibinfo{author}{Movshon, J.}
\newblock \bibinfo{title}{A computational analysis of the relationship between
  neuronal and behavioral responses to visual motion}.
\newblock \emph{\bibinfo{journal}{The Journal of Neuroscience}}
  \textbf{\bibinfo{volume}{16}}, \bibinfo{pages}{1486--1510}
  (\bibinfo{year}{1996}).

\bibitem{Seung1993}
\bibinfo{author}{Seung, H.} \& \bibinfo{author}{Sompolinsky, H.}
\newblock \bibinfo{title}{Simple models for reading neuronal population codes}.
\newblock \emph{\bibinfo{journal}{Proceedings of the National Academy of
  Sciences}} \textbf{\bibinfo{volume}{90}}, \bibinfo{pages}{10749--10753}
  (\bibinfo{year}{1993}).

\bibitem{Zemel1998}
\bibinfo{author}{Zemel, R.}, \bibinfo{author}{Dayan, P.} \&
  \bibinfo{author}{Pouget, A.}
\newblock \bibinfo{title}{Probabilistic interpretation of population codes.}
\newblock \emph{\bibinfo{journal}{Neural Computation}}
  \textbf{\bibinfo{volume}{10}}, \bibinfo{pages}{403--430}
  (\bibinfo{year}{1998}).

\bibitem{Zhang1999}
\bibinfo{author}{Zhang, K.} \& \bibinfo{author}{Sejnowski, T.}
\newblock \bibinfo{title}{Neuronal tuning: To sharpen or broaden?}
\newblock \emph{\bibinfo{journal}{Neural Computation}}
  \textbf{\bibinfo{volume}{11}}, \bibinfo{pages}{75--84}
  (\bibinfo{year}{1999}).

\bibitem{Pouget1999}
\bibinfo{author}{Pouget, A.}, \bibinfo{author}{Deneve, S.},
  \bibinfo{author}{Ducom, J.} \& \bibinfo{author}{Latham, P.}
\newblock \bibinfo{title}{Narrow versus wide tuning curves: What's best for a
  population code?}
\newblock \emph{\bibinfo{journal}{Neural Computation}}
  \textbf{\bibinfo{volume}{11}}, \bibinfo{pages}{85--90}
  (\bibinfo{year}{1999}).

\bibitem{Dayan2001}
\bibinfo{author}{Dayan, P.} \& \bibinfo{author}{Abbott, L.}
\newblock \emph{\bibinfo{title}{Theoretical Neuroscience: Computational and
  Mathematical Modeling of Neural Systems}} (\bibinfo{publisher}{The {MIT}
  Press}, \bibinfo{year}{2001}).
\newblock \bibinfo{note}{Published: Hardcover}.

\bibitem{Ganguli10c}
\bibinfo{author}{Ganguli, D.} \& \bibinfo{author}{Simoncelli, E.~P.}
\newblock \bibinfo{title}{Implicit encoding of prior probabilities in optimal
  neural populations}.
\newblock In \bibinfo{editor}{Lafferty, J.}, \bibinfo{editor}{Williams, C.
  K.~I.}, \bibinfo{editor}{Shawe-Taylor, J.}, \bibinfo{editor}{Zemel, R.} \&
  \bibinfo{editor}{Culotta, A.} (eds.) \emph{\bibinfo{booktitle}{Adv. Neural
  Information Processing Systems (NIPS*10)}}, vol.~\bibinfo{volume}{23},
  \bibinfo{pages}{658--666} (\bibinfo{publisher}{MIT Press},
  \bibinfo{address}{Cambridge, MA}, \bibinfo{year}{2010}).
\newblock \bibinfo{note}{Presented at Neural Information Processing Systems 23,
  Vancouver, Dec 6-9 2010.}

\bibitem{Ganguli12b}
\bibinfo{author}{Ganguli, D.} \& \bibinfo{author}{Simoncelli, E.~P.}
\newblock \bibinfo{title}{Implicit embedding of prior probabilities in
  optimally efficient neural populations}.
\newblock \bibinfo{type}{Tech. Rep.} \bibinfo{number}{1209.5006},
  \bibinfo{institution}{ArXiv e-prints (arXiv.org)} (\bibinfo{year}{2012}).
\newblock \urlprefix\url{http://arxiv.org/abs/1209.5006}.

\bibitem{Ma2006}
\bibinfo{author}{Ma, W.}, \bibinfo{author}{Beck, J.}, \bibinfo{author}{Latham,
  P.} \& \bibinfo{author}{Pouget, A.}
\newblock \bibinfo{title}{Bayesian inference with probabilistic population
  codes}.
\newblock \emph{\bibinfo{journal}{Nature Neuroscience}}
  \textbf{\bibinfo{volume}{9}}, \bibinfo{pages}{1432--1438}
  (\bibinfo{year}{2006}).

\bibitem{Lennie03}
\bibinfo{author}{Lennie, P.}
\newblock \bibinfo{title}{The cost of cortical computation}.
\newblock \emph{\bibinfo{journal}{Current Biology}}
  \textbf{\bibinfo{volume}{13}}, \bibinfo{pages}{493--497}
  (\bibinfo{year}{2003}).

\bibitem{Laughlin98}
\bibinfo{author}{Laughlin, S.~B.}, \bibinfo{author}{de~Ruyter~van Steveninck,
  R.} \& \bibinfo{author}{Anderson, J.~C.}
\newblock \bibinfo{title}{The metabolic cost of information}.
\newblock \emph{\bibinfo{journal}{Nature neuroscience}}
  \textbf{\bibinfo{volume}{1}}, \bibinfo{pages}{36--41} (\bibinfo{year}{1998}).

\bibitem{Cox1974}
\bibinfo{author}{Cox, D.} \& \bibinfo{author}{Hinkley, D.}
\newblock \emph{\bibinfo{title}{Theoretical statistics}}
  (\bibinfo{publisher}{London: Chapman and Hall.}, \bibinfo{year}{1974}).

\bibitem{Series2009}
\bibinfo{author}{Seri\`{e}s, P.}, \bibinfo{author}{Stocker, A.} \&
  \bibinfo{author}{{EP}, S.}
\newblock \bibinfo{title}{Is the homunculus
  {\textquotedblleft}aware{\textquotedblright} of sensory adaptation?}
\newblock \emph{\bibinfo{journal}{Neural Computation}}
  \textbf{\bibinfo{volume}{21}}, \bibinfo{pages}{3271--3304}
  (\bibinfo{year}{2009}).

\bibitem{Brunel1998}
\bibinfo{author}{Brunel, N.} \& \bibinfo{author}{Nadal, J.}
\newblock \bibinfo{title}{Mutual information, fisher information, and
  population coding}.
\newblock \emph{\bibinfo{journal}{Neural Computation}}
  \textbf{\bibinfo{volume}{10}}, \bibinfo{pages}{1731--1757}
  (\bibinfo{year}{1998}).

\bibitem{Storm1994}
\bibinfo{author}{Storm, J.}
\newblock \bibinfo{title}{Great smokey mountains national park: winter and
  spring. in: The macaulay library of natural sounds. ithaca, {NY:} cornell
  laboratory of ornithology.} (\bibinfo{year}{1994}).

\bibitem{Storm1994a}
\bibinfo{author}{Storm, J.}
\newblock \bibinfo{title}{Great smokey mountains national park: summer and
  fall. in: The macaulay library of natural sounds. ithaca, {NY:} cornell
  laboratory of ornithology.} (\bibinfo{year}{1994}).

\bibitem{Emmons1997}
\bibinfo{author}{Emmons, L.}, \bibinfo{author}{Whitney, B.} \&
  \bibinfo{author}{Ross, D.}
\newblock \bibinfo{title}{Sounds of neotropical rainforrest mammals. in: The
  macaulay library of natural sounds. ithaca, {NY:} cornell laboratory of
  ornithology.} (\bibinfo{year}{1997}).

\bibitem{Hateren1998}
\bibinfo{author}{van Hateren, J.} \& \bibinfo{author}{van~der Schaaf, A.}
\newblock \bibinfo{title}{Independent component filters of natural images
  compared with simple cells in primary visual cortex.}
\newblock \emph{\bibinfo{journal}{Proceedings of the Royal Society B:
  Biological Sciences}} \textbf{\bibinfo{volume}{265}},
  \bibinfo{pages}{359--366} (\bibinfo{year}{1998}).

\bibitem{Doi2003}
\bibinfo{author}{Doi, E.}, \bibinfo{author}{Inui, T.}, \bibinfo{author}{Lee,
  T.}, \bibinfo{author}{Wachtler, T.} \& \bibinfo{author}{Sejnowski, T.}
\newblock \bibinfo{title}{Spatiochromatic receptive field properties derived
  from information-theoretic analyses of cone mosaic responses to natural
  scenes.}
\newblock \emph{\bibinfo{journal}{Neural Comput}}
  \textbf{\bibinfo{volume}{15}}, \bibinfo{pages}{397--417}
  (\bibinfo{year}{2003}).

\bibitem{Olmos2004}
\bibinfo{author}{Olmos, A.} \& \bibinfo{author}{Kingdom, F.}
\newblock \emph{\bibinfo{title}{{McGill} Calibrated Image Database,
  http://tabby.vision.mcgill.ca}} (\bibinfo{year}{2004}).

\bibitem{Carney1999}
\bibinfo{author}{Carney, L.}, \bibinfo{author}{{McDuffy}, M.} \&
  \bibinfo{author}{Shekhter, I.}
\newblock \bibinfo{title}{Frequency glides in the impulse responses of
  auditory-nerve fibers}.
\newblock \emph{\bibinfo{journal}{The Journal of the Acoustical Society of
  America}} \textbf{\bibinfo{volume}{105}}, \bibinfo{pages}{2384}
  (\bibinfo{year}{1999}).

\bibitem{Smith2006}
\bibinfo{author}{Smith, E.} \& \bibinfo{author}{Lewicki, M.}
\newblock \bibinfo{title}{Efficient auditory coding}.
\newblock \emph{\bibinfo{journal}{Nature}} \textbf{\bibinfo{volume}{439}},
  \bibinfo{pages}{978--982} (\bibinfo{year}{2006}).

\bibitem{Rodriguez2010}
\bibinfo{author}{Rodr\'{i}guez, F.}, \bibinfo{author}{Chen, C.},
  \bibinfo{author}{Read, H.} \& \bibinfo{author}{{MA}, E.}
\newblock \bibinfo{title}{Neural modulation tuning characteristics scale to
  efficiently encode natural sound statistics}.
\newblock \emph{\bibinfo{journal}{The Journal of Neuroscience}}
  \textbf{\bibinfo{volume}{30}}, \bibinfo{pages}{15969 --15980}
  (\bibinfo{year}{2010}).

\bibitem{Mansfield1974}
\bibinfo{author}{Mansfield, R.}
\newblock \bibinfo{title}{Neural basis of orientation perception in primate
  vision}.
\newblock \emph{\bibinfo{journal}{Science}} \textbf{\bibinfo{volume}{186}},
  \bibinfo{pages}{1133 --1135} (\bibinfo{year}{1974}).

\bibitem{Cavanaugh2002}
\bibinfo{author}{Cavanaugh, J.}, \bibinfo{author}{Bair, W.} \&
  \bibinfo{author}{Movshon, J.}
\newblock \bibinfo{title}{Nature and interaction of signals from the receptive
  field center and surround in macaque v1 neurons}.
\newblock \emph{\bibinfo{journal}{Journal of Neurophysiology}}
  \textbf{\bibinfo{volume}{88}}, \bibinfo{pages}{2530 --2546}
  (\bibinfo{year}{2002}).

\bibitem{Movshon15}
\bibinfo{author}{Wang, H.} \& \bibinfo{author}{Movshon, J.}
\newblock \bibinfo{title}{Properties of pattern and component direction
  selective cells in area {MT}of the macaque}.
\newblock \emph{\bibinfo{journal}{J Neurophysiology}}  (\bibinfo{year}{2015}).

\bibitem{Moore1973}
\bibinfo{author}{Moore, B.}
\newblock \bibinfo{title}{Frequency difference limens for short-duration
  tones}.
\newblock \emph{\bibinfo{journal}{The Journal of the Acoustical Society of
  America}} \textbf{\bibinfo{volume}{54}}, \bibinfo{pages}{610}
  (\bibinfo{year}{1973}).

\bibitem{Wier1977}
\bibinfo{author}{Wier, C.}
\newblock \bibinfo{title}{Frequency discrimination as a function of frequency
  and sensation level}.
\newblock \emph{\bibinfo{journal}{The Journal of the Acoustical Society of
  America}} \textbf{\bibinfo{volume}{61}}, \bibinfo{pages}{178}
  (\bibinfo{year}{1977}).

\bibitem{Formby1985}
\bibinfo{author}{Formby, C.}
\newblock \bibinfo{title}{Differential sensitivity to tonal frequency and to
  the rate of amplitude modulation of broadband noise by normally hearing
  listeners}.
\newblock \emph{\bibinfo{journal}{The Journal of the Acoustical Society of
  America}} \textbf{\bibinfo{volume}{78}}, \bibinfo{pages}{70}
  (\bibinfo{year}{1985}).

\bibitem{Lemanska2002}
\bibinfo{author}{Lemanska, J.}, \bibinfo{author}{Sek, A.} \&
  \bibinfo{author}{{EB}, S.}
\newblock \bibinfo{title}{Discrimination of the amplitude modulation rate}.
\newblock \emph{\bibinfo{journal}{Archives of Acoustics}}
  \textbf{\bibinfo{volume}{27}}, \bibinfo{pages}{3--21} (\bibinfo{year}{2002}).

\bibitem{Girshick2011}
\bibinfo{author}{Girshick, A.}, \bibinfo{author}{Landy, M.} \&
  \bibinfo{author}{Simoncelli, E.}
\newblock \bibinfo{title}{Cardinal rules: visual orientation perception
  reflects knowledge of environmental statistics}.
\newblock \emph{\bibinfo{journal}{Nature Neuroscience}}
  \textbf{\bibinfo{volume}{14}}, \bibinfo{pages}{926--932}
  (\bibinfo{year}{2011}).

\bibitem{Caeli1983}
\bibinfo{author}{Caeli, T.}, \bibinfo{author}{Brettel, H.},
  \bibinfo{author}{Rentschler, I.} \& \bibinfo{author}{Hilz, R.}
\newblock \bibinfo{title}{Discrimination thresholds in the two-dimensional
  spatial frequency domain}.
\newblock \emph{\bibinfo{journal}{Vision Research}}
  \textbf{\bibinfo{volume}{23}}, \bibinfo{pages}{129--133}
  (\bibinfo{year}{1983}).

\bibitem{Regan1982}
\bibinfo{author}{Regan, D.}, \bibinfo{author}{Bartol, S.},
  \bibinfo{author}{Beverly, T.} \& \bibinfo{author}{Murray, T.}
\newblock \bibinfo{title}{Spatial frequency discrimination in normal vision and
  in patients with multiple sclerosis}.
\newblock \emph{\bibinfo{journal}{Brain}} \textbf{\bibinfo{volume}{105}},
  \bibinfo{pages}{735 --754} (\bibinfo{year}{1982}).

\bibitem{McKee1984}
\bibinfo{author}{{McKee}, S.} \& \bibinfo{author}{Nakayama, K.}
\newblock \bibinfo{title}{The detection of motion in the peripheral visual
  field}.
\newblock \emph{\bibinfo{journal}{Vision Research}}
  \textbf{\bibinfo{volume}{24}}, \bibinfo{pages}{25--32}
  (\bibinfo{year}{1984}).

\bibitem{DeBruyn1988}
\bibinfo{author}{De~Bruyn, B.} \& \bibinfo{author}{Orban, G.}
\newblock \bibinfo{title}{Human velocity and direction discrimination measured
  with random dot patterns}.
\newblock \emph{\bibinfo{journal}{Vision Research}}
  \textbf{\bibinfo{volume}{28}}, \bibinfo{pages}{1323--1335}
  (\bibinfo{year}{1988}).

\bibitem{Nover2005}
\bibinfo{author}{Nover, H.}, \bibinfo{author}{Anderson, C.} \&
  \bibinfo{author}{{DeAngelis}, G.}
\newblock \bibinfo{title}{A logarithmic, {Scale-Invariant} representation of
  speed in macaque middle temporal area accounts for speed discrimination
  performance}.
\newblock \emph{\bibinfo{journal}{The Journal of Neuroscience}}
  \textbf{\bibinfo{volume}{25}}, \bibinfo{pages}{10049--10060}
  (\bibinfo{year}{2005}).

\bibitem{Harper2004}
\bibinfo{author}{Harper, N.} \& \bibinfo{author}{{McAlpine}, D.}
\newblock \bibinfo{title}{Optimal neural population coding of an auditory
  spatial cue}.
\newblock \emph{\bibinfo{journal}{Nature}} \textbf{\bibinfo{volume}{430}},
  \bibinfo{pages}{682--686} (\bibinfo{year}{2004}).

\bibitem{Laughlin1981}
\bibinfo{author}{Laughlin, S.}
\newblock \bibinfo{title}{A simple coding procedure enhances a neuron's
  information capacity}.
\newblock \emph{\bibinfo{journal}{Zeitschrift F\"{u}r Naturforschung.}}
  \textbf{\bibinfo{volume}{36}}, \bibinfo{pages}{910--912}
  (\bibinfo{year}{1981}).

\bibitem{Nadal1994}
\bibinfo{author}{Nadal, J.} \& \bibinfo{author}{Parga, N.}
\newblock \bibinfo{title}{Non linear neurons in the low noise limit: A
  factorial code maximizes information transfer}.
\newblock \emph{\bibinfo{journal}{Network: Computation in Neural Systems}}
  (\bibinfo{year}{1994}).

\bibitem{McDonnell2008}
\bibinfo{author}{{McDonnell}, M.} \& \bibinfo{author}{Stocks, N.}
\newblock \bibinfo{title}{Maximally informative stimuli and tuning curves for
  sigmoidal rate-coding neurons and populations}.
\newblock \emph{\bibinfo{journal}{Physical Review Letters}}
  \textbf{\bibinfo{volume}{101}}, \bibinfo{pages}{58103}
  (\bibinfo{year}{2008}).

\bibitem{Olshausen1996}
\bibinfo{author}{Olshausen, B.} \& \bibinfo{author}{Field, D.}
\newblock \bibinfo{title}{Emergence of simple-cell receptive field properties
  by learning a sparse code for natural images}.
\newblock \emph{\bibinfo{journal}{Nature}} \textbf{\bibinfo{volume}{381}},
  \bibinfo{pages}{607--609} (\bibinfo{year}{1996}).

\bibitem{Bell1997}
\bibinfo{author}{Bell, A.} \& \bibinfo{author}{Sejnowski, T.}
\newblock \bibinfo{title}{The "independent components" of natural scenes are
  edge filters}.
\newblock \emph{\bibinfo{journal}{Vision Research}}
  \textbf{\bibinfo{volume}{37}}, \bibinfo{pages}{3327--3338}
  (\bibinfo{year}{1997}).

\bibitem{Geisler2009}
\bibinfo{author}{Geisler, W.}, \bibinfo{author}{Najemnik, J.} \&
  \bibinfo{author}{Ing, A.}
\newblock \bibinfo{title}{Optimal stimulus encoders for natural tasks}.
\newblock \emph{\bibinfo{journal}{Journal of Vision}}
  \textbf{\bibinfo{volume}{9}}, \bibinfo{pages}{17.1--16}
  (\bibinfo{year}{2009}).

\bibitem{Twer2001}
\bibinfo{author}{von~der Twer, T.} \& \bibinfo{author}{{MacLeod}, D.}
\newblock \bibinfo{title}{Optimal nonlinear codes for the perception of natural
  colours.}
\newblock \emph{\bibinfo{journal}{Network}} \textbf{\bibinfo{volume}{12}},
  \bibinfo{pages}{395--407} (\bibinfo{year}{2001}).

\bibitem{Helmholtz2000}
\bibinfo{author}{Helmholtz, H.}
\newblock \emph{\bibinfo{title}{Treatise on physiological optics}}
  (\bibinfo{publisher}{Thoemmes Press}, \bibinfo{address}{Bristol, {UK}},
  \bibinfo{year}{2000}).

\bibitem{Jazayeri2006}
\bibinfo{author}{Jazayeri, M.} \& \bibinfo{author}{Movshon, J.}
\newblock \bibinfo{title}{Optimal representation of sensory information by
  neural populations}.
\newblock \emph{\bibinfo{journal}{Nature Neuroscience}}
  \textbf{\bibinfo{volume}{9}}, \bibinfo{pages}{690--696}
  (\bibinfo{year}{2006}).

\bibitem{Knill1996}
\bibinfo{author}{Knill, D.} \& \bibinfo{author}{Richards, W.}
\newblock \emph{\bibinfo{title}{Perception as Bayesian inference}}
  (\bibinfo{publisher}{Cambridge University Press},
  \bibinfo{address}{Cambridge, {UK}}, \bibinfo{year}{1996}).

\bibitem{Simoncelli2009}
\bibinfo{author}{Simoncelli, E.}
\newblock \bibinfo{title}{Optimal estimation in sensory systems}.
\newblock In \bibinfo{editor}{Gazzaniga, M.} (ed.)
  \emph{\bibinfo{booktitle}{The Cognitive Neurosciences, {IV}}},
  \bibinfo{pages}{525--535} (\bibinfo{publisher}{{MIT} Press},
  \bibinfo{year}{2009}).

\bibitem{Ganguli12a}
\bibinfo{author}{Ganguli, D.} \& \bibinfo{author}{Simoncelli, E.~P.}
\newblock \bibinfo{title}{Neural implementation of {Bayesian} inference using
  efficient population codes}.
\newblock In \emph{\bibinfo{booktitle}{Computational and Systems Neuroscience
  (CoSyNe)}}, \bibinfo{number}{II-9} (\bibinfo{address}{Salt Lake City, Utah},
  \bibinfo{year}{2012}).

\bibitem{Lennie2003}
\bibinfo{author}{Lennie, P.}
\newblock \bibinfo{title}{The cost of cortical computation}.
\newblock \emph{\bibinfo{journal}{Current Biology}}
  \textbf{\bibinfo{volume}{13}}, \bibinfo{pages}{493--497}
  (\bibinfo{year}{2003}).

\bibitem{Yarrow2012}
\bibinfo{author}{Yarrow, S.}, \bibinfo{author}{Challis, E.} \&
  \bibinfo{author}{Seri\`{e}s, P.}
\newblock \bibinfo{title}{Fisher and shannon information in finite neural
  populations}.
\newblock \emph{\bibinfo{journal}{Neural Computation}}
  \textbf{\bibinfo{volume}{24}}, \bibinfo{pages}{1740--1780}
  (\bibinfo{year}{2012}).

\bibitem{Welch1967}
\bibinfo{author}{Welch, P.}
\newblock \bibinfo{title}{The use of fast fourier transform for the estimation
  of power spectra: A method based on time averaging over short, modified
  periodograms}.
\newblock \emph{\bibinfo{journal}{Audio and Electroacoustics, {IEEE}
  Transactions on}} \textbf{\bibinfo{volume}{15}}, \bibinfo{pages}{70 -- 73}
  (\bibinfo{year}{1967}).

\bibitem{Attias1997}
\bibinfo{author}{Attias, H.} \& \bibinfo{author}{Schreiner, C.}
\newblock \bibinfo{title}{Temporal {Low-Order} statistics of natural sounds}.
\newblock \emph{\bibinfo{journal}{{NIPS}}} \textbf{\bibinfo{volume}{9}},
  \bibinfo{pages}{27---33} (\bibinfo{year}{1997}).

\bibitem{McDermott2011}
\bibinfo{author}{{McDermott}, J.} \& \bibinfo{author}{Simoncelli, E.}
\newblock \bibinfo{title}{Sound texture perception via statistics of the
  auditory periphery: Evidence from sound synthesis}.
\newblock \emph{\bibinfo{journal}{Neuron}} \textbf{\bibinfo{volume}{71}},
  \bibinfo{pages}{926--940} (\bibinfo{year}{2011}).

\bibitem{Glasberg1990}
\bibinfo{author}{Glasberg, B.} \& \bibinfo{author}{Moore, B.}
\newblock \bibinfo{title}{Derivation of auditory filter shapes from
  notched-noise data}.
\newblock \emph{\bibinfo{journal}{Hearing Research}}
  \textbf{\bibinfo{volume}{47}}, \bibinfo{pages}{103--138}
  (\bibinfo{year}{1990}).

\bibitem{Burt1983}
\bibinfo{author}{Burt, P.} \& \bibinfo{author}{Adelson, E.}
\newblock \bibinfo{title}{The laplacian pyramid as a compact image code}.
\newblock \emph{\bibinfo{journal}{Communications, {IEEE} Transactions on}}
  \textbf{\bibinfo{volume}{31}}, \bibinfo{pages}{532 -- 540}
  (\bibinfo{year}{1983}).

\bibitem{Farid2004}
\bibinfo{author}{Farid, H.} \& \bibinfo{author}{Simoncelli, E.}
\newblock \bibinfo{title}{Differentiation of discrete multidimensional
  signals}.
\newblock \emph{\bibinfo{journal}{Image Processing, {IEEE} Transactions on}}
  \textbf{\bibinfo{volume}{13}}, \bibinfo{pages}{496 --508}
  (\bibinfo{year}{2004}).

\bibitem{Granlund1995}
\bibinfo{author}{Granlund, G.} \& \bibinfo{author}{Knutsson, H.}
\newblock \emph{\bibinfo{title}{Signal Processing for Computer Vision}}
  (\bibinfo{publisher}{Kluwer Academic Publishers}, \bibinfo{address}{Norwell,
  Massachusetts}, \bibinfo{year}{1995}).

\bibitem{Field1987}
\bibinfo{author}{Field, D.}
\newblock \bibinfo{title}{Relations between the statistics of natural images
  and the response properties of cortical cells}.
\newblock \emph{\bibinfo{journal}{Journal of the Optical Society of America A}}
  \textbf{\bibinfo{volume}{4}}, \bibinfo{pages}{2379--2394}
  (\bibinfo{year}{1987}).

\bibitem{Ruderman1994}
\bibinfo{author}{Ruderman, D.} \& \bibinfo{author}{Bialek, W.}
\newblock \bibinfo{title}{Statistics of natural images: Scaling in the woods}.
\newblock \emph{\bibinfo{journal}{Physical Review Letters}}
  \textbf{\bibinfo{volume}{73}}, \bibinfo{pages}{814--817}
  (\bibinfo{year}{1994}).

\bibitem{Dong1995}
\bibinfo{author}{Dong, D.} \& \bibinfo{author}{Atick, J.}
\newblock \bibinfo{title}{Statistics of natural {Time-Varying} images}.
\newblock \emph{\bibinfo{journal}{Network: Computation in Neural Systems}}
  \textbf{\bibinfo{volume}{6}}, \bibinfo{pages}{345---358}
  (\bibinfo{year}{1995}).

\bibitem{Roth2007}
\bibinfo{author}{Roth, S.} \& \bibinfo{author}{Black, M.}
\newblock \bibinfo{title}{On the spatial statistics of optical flow}.
\newblock \emph{\bibinfo{journal}{International Journal of Computer Vision}}
  \textbf{\bibinfo{volume}{74}}, \bibinfo{pages}{33--50}
  (\bibinfo{year}{2007}).

\bibitem{Stocker2006a}
\bibinfo{author}{Stocker, A.} \& \bibinfo{author}{Simoncelli, E.}
\newblock \bibinfo{title}{Noise characteristics and prior expectations in human
  visual speed perception}.
\newblock \emph{\bibinfo{journal}{Nature Neuroscience}}
  \textbf{\bibinfo{volume}{9}}, \bibinfo{pages}{578--585}
  (\bibinfo{year}{2006}).

\bibitem{Baowang2003}
\bibinfo{author}{Baowang, L.}, \bibinfo{author}{Peterson, M.} \&
  \bibinfo{author}{Freeman, R.}
\newblock \bibinfo{title}{Oblique effect: A neural basis in the visual cortex}.
\newblock \emph{\bibinfo{journal}{Journal of Neurophysiology}}
  \textbf{\bibinfo{volume}{90}}, \bibinfo{pages}{204 --217}
  (\bibinfo{year}{2003}).

\bibitem{Appelle1972}
\bibinfo{author}{Appelle, S.}
\newblock \bibinfo{title}{Perception and discrimination as a function of
  stimulus orientation: The "oblique effect" in man and animals}.
\newblock \emph{\bibinfo{journal}{Psychological Bulletin}}
  \textbf{\bibinfo{volume}{78}}, \bibinfo{pages}{266--278}
  (\bibinfo{year}{1972}).

\bibitem{Rothkopf2009}
\bibinfo{author}{Rothkopf, C.}, \bibinfo{author}{Weisswange, T.} \&
  \bibinfo{author}{Triesch, J.}
\newblock \bibinfo{title}{Learning independent causes in natural images
  explains the spacevariant oblique effect}.
\newblock In \emph{\bibinfo{booktitle}{Development and Learning, 2009. {ICDL}
  2009. {IEEE} 8th International Conference on}}, \bibinfo{pages}{1 --6}
  (\bibinfo{year}{2009}).

\bibitem{Freeman2011}
\bibinfo{author}{Freeman, J.}, \bibinfo{author}{Brouwer, G.},
  \bibinfo{author}{Heeger, D.} \& \bibinfo{author}{Merriam, E.}
\newblock \bibinfo{title}{Orientation decoding depends on maps, not columns}.
\newblock \emph{\bibinfo{journal}{The Journal of Neuroscience}}
  \textbf{\bibinfo{volume}{31}}, \bibinfo{pages}{4792--4804}
  (\bibinfo{year}{2011}).

\bibitem{Hyvarinen1999}
\bibinfo{author}{Hyvarinen, A.}
\newblock \bibinfo{title}{Fast and robust fixed-point algorithms for
  independent component analysis}.
\newblock \emph{\bibinfo{journal}{Neural Networks, {IEEE} Transactions on}}
  \textbf{\bibinfo{volume}{10}}, \bibinfo{pages}{626 --634}
  (\bibinfo{year}{1999}).

\bibitem{Ganguli2010}
\bibinfo{author}{Ganguli, D.} \& \bibinfo{author}{Simoncelli, E.}
\newblock \bibinfo{title}{Implicit encoding of prior probabilities in optimal
  neural populations}.
\newblock In \bibinfo{editor}{Lafferty, J.}, \bibinfo{editor}{Williams, C.},
  \bibinfo{editor}{{Shawe-Taylor}, J.}, \bibinfo{editor}{Zemel, R.} \&
  \bibinfo{editor}{Culotta, A.} (eds.) \emph{\bibinfo{booktitle}{Advances in
  Neural Information Processing Systems {(NIPS)}}}, vol.~\bibinfo{volume}{23},
  \bibinfo{pages}{658--666} (\bibinfo{address}{Salt Lake City, Utah},
  \bibinfo{year}{2010}).

\end{thebibliography}

\bibliographystyle{naturemag}

\newpage
\section*{Appendix} 

\subsection*{Statistical test for cell density estimates} 

To quantify the accuracy of our prediction for cell density, we compared the cumulative density function (CDF) of the environmental distribution, and the empirical CDF of the cell density \figref{A1}. For all attributes, we find that the data are much closer to our predictions than a uniform distribution (Fig.~A1, grey lines). As can be seen here (and in the density plots of \figref{2k-o}), empirically measured cell densities deviate systematically from the predictions. This may be attributable to inherent biases in experimental procedures: In single-cell electrophysiology experiments, the reported cells are typically found and isolated by advancing the electrode while delivering stimuli, and those stimuli tend to be drawn from the center of some typical parameter range. We quantified the accuracy of the predictions with a single sample Kolmogorov-Smirnov (KS) test, which computes the maximal difference between the empirical and predicted CDFs. For the attribute of retinal speed, we cannot reject the null hypothesis that the cell population is drawn from the estimated environmental distribution ($p = 0.15$, KS test). For acoustic frequency, modulation frequency, and spatial frequency, the cell densities deviated more significantly from the corresponding environmental distributions ($p < 0.001$ for each attribute, KS test).

\begin{figure}[h]
\begin{center}
{
\includegraphics[scale=.91]{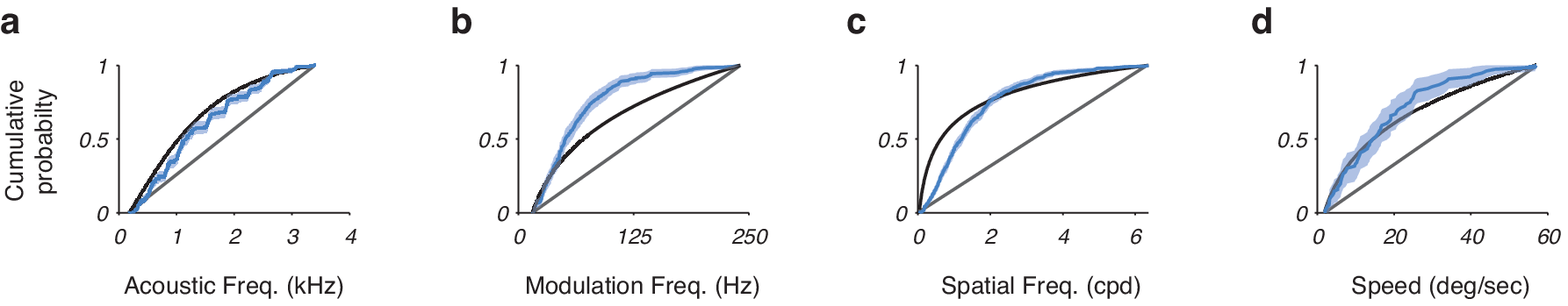}} 
\end{center}
\caption*{{\bf Fig A1. Cumulative cell density comparisons.} Cumulative distribution of the optimally efficient cell density (black line), a homogeneous cell density (grey diagonal line), and that of the measured physiological population (blue line). Blue shaded regions denote $95\%$ confidence intervals of the empirical cumulative distributions, obtained from $1,000$ bootstrap samples. {\bf a}, Acoustic frequency. {\bf b}, Modulation frequency. {\bf c}, Spatial frequency. {\bf d}, Speed.}\label{supp-kstest} 
\end{figure}

\subsection*{Comparison of gain predictions to physiological data} 

\begin{figure}
\begin{center}
{
\includegraphics[scale=1.3]{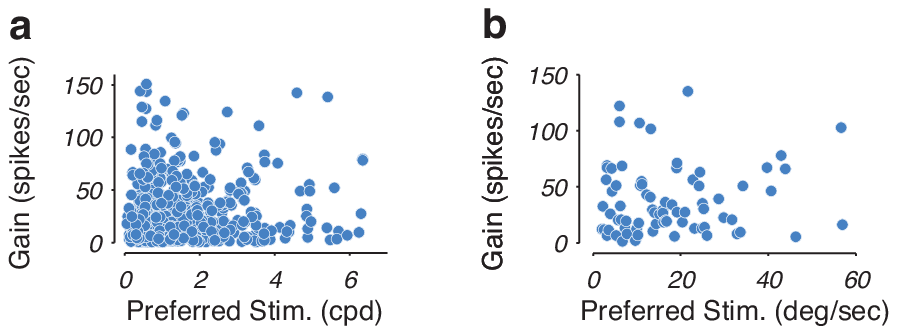}} 
\end{center}
\caption*{{\bf Fig A2. Relationship of response maximum (gain) to preferred stimulus value.} Cell gain (measured as the maximum average firing rate of each cell) plotted as a function of preferred stimulus value. {\bf a}, Data from $538$ cells in Macaque V1 (same data shown in Fig.~2i,n). {\bf b}, Data from $76$ Macaque MT cells (same data shown in Fig.~2j,o).} \label{supp-gain} 
\end{figure}

Our framework predicts that the tuning curves of cells in an optimal population should all have the same maximum firing rate (``gain''). We examined this prediction in the V1 spatial frequency\cite{Cavanaugh2002}, and MT speed\cite{Movshon15} data sets \figref{A2}, as well as an orientation-tuned population recorded in cat area $17$ \cite{Baowang2003} \figref{A3c}. Although we find significant variability in gain values, it is not systematically related to the stimulus values ($r = 0.065, p = 0.13$ for spatial frequency and $r = 0.104, p = 0.37$ for speed).

To test for possible nonlinear relationships between the gain, $g$, and preferred stimuli, $\mu$, we constructed null model in which the two quantities are statistically independent: $p(\mu, g) = p(\mu)p(g)$. The distribution for $p(\mu)$ was fit with an exponential distribution with mean parameter $1.53 \text{ cpd}$ for spatial frequency and $17.35 \text{ deg/sec}$ for speed, chosen to maximize the log likelihood computed over the data. The distribution for $p(g)$ was also fit with an exponential distribution with maximum likelihood parameters of $27.01 \text{ spikes/sec}$ for spatial frequency and $29.71 \text{ spikes/sec}$ for speed. For each attribute, we generated samples of preferred stimuli and gain from the fitted null model, matched to the sample size of the data. We computed the log likelihood for $10,000$ of these simulated data sets, and then compared the log likelihood of the actual data set to this distribution. For each attribute, we found that the log likelihood of the data under the independent model was well within the $95\%$ confidence intervals computed from the distribution of log likelihoods of synthetic data sampled from the model. Therefore we cannot reject the hypothesis that the gain and preferred stimuli of the cells are statistically independent.

\subsection*{Another physiological data set for orientation tuning} 
Visually oriented stimuli are more perceptually discriminable about horizontal and vertical axes than oblique axes. This ``oblique effect'' has been confirmed empirically in numerous behavioral studies in humans and animals published over the past century\cite{Appelle1972}. Our efficient coding framework makes strong predictions about the magnitude of the oblique effect: orientation discrimination thresholds should be inversely proportional to the frequency of occurrence of orientations in the natural environment. We found that this prediction is highly accurate \figref{2r}. However, it relies on the intermediate physiological prediction, that more frequently occurring orientations should be coded with more cells, and that those cells should have narrower tuning. Although we have shown that this holds for a data set drawn from near-foveal macaque V1 \figref{2m}, other physiological investigations into the anisotropy of orientation tuning preferences have yielded mixed results \cite{Baowang2003}.

\begin{figure}
\begin{center}
{
\includegraphics[scale=1.2]{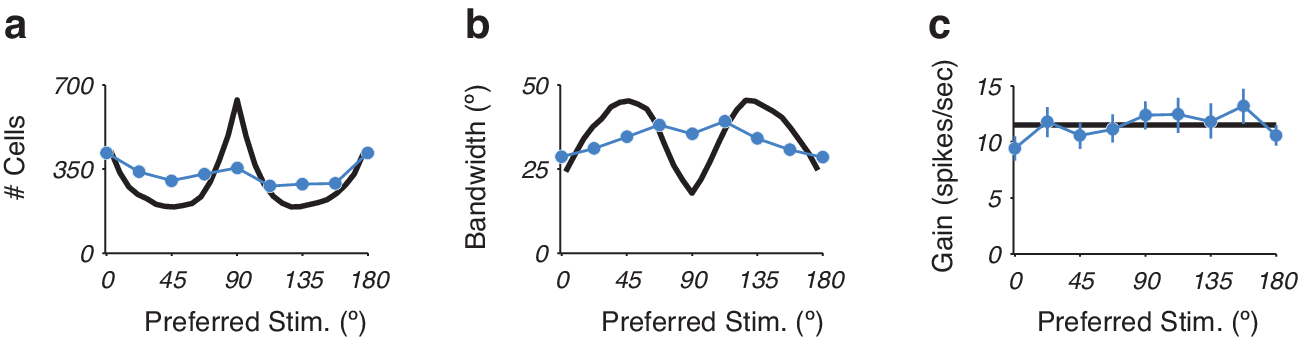}} 
\end{center}
\label{supp-cat} \caption*{{\bf Fig A3. Orientation tuning in cat area 17.} Orientation tuning preferences of $2,598$ simple cells in cat area $17$\cite{Baowang2003}, and predictions derived from the environmental distribution (black curves). {\bf a}, Number of cells tuned to each orientation. The distribution shows clear peaks at horizontal and vertical orientations that are statistically significant ($p < 0.05, \chi^2 \text{ test}$). {\bf b}, Orientation tuning width as a function of preferred orientation. The cells show a significant narrowing of orientation tuning at horizontal orientations, but not to vertical orientations. {\bf c}, Mean response amplitude (gain) as a function of preferred orientation.} 
\end{figure}

We analyzed a second data set containing measurements of orientation tuning in simple cells, recorded from cat area $17$\cite{Baowang2003} \figref{A3}. There are again more cells with narrower tuning for more frequently occurring orientations. However, the magnitudes of these biases are significantly smaller than those found in the Macaque V1 data \figref{2m} and are not well-matched to the predictions of our theory that come from either the environmental or perceptual data (Fig.~A3, thick black lines). There are several possible reasons for these inconsistencies. First, the measurements are made in different species, and may reflect differences in either the neurophysiology, or perhaps the corresponding visual environments, of cats and monkeys. Second, the cells in the Macaque data set had receptive fields near the fovea. In the same study, it was found that the density of cells recorded in the periphery was nearly uniform\cite{Mansfield1974}. The cells in the cat data set, on the other hand, had receptive fields at a wide range of eccentricities, potentially diminishing the magnitude of the reported heterogeneity in cell density. Finally, there is theoretical and physiological evidence that orientation tuning biases may also vary as a function of the angular position receptive fields relative to the fovea\cite{Rothkopf2009,Freeman2011}. A resolution of these discrepancies await a large-scale study that can disentangle these effects.

\subsection*{Comparison to alternative efficient coding frameworks}

\begin{figure}
\begin{center}
{
\includegraphics[scale=1.2]{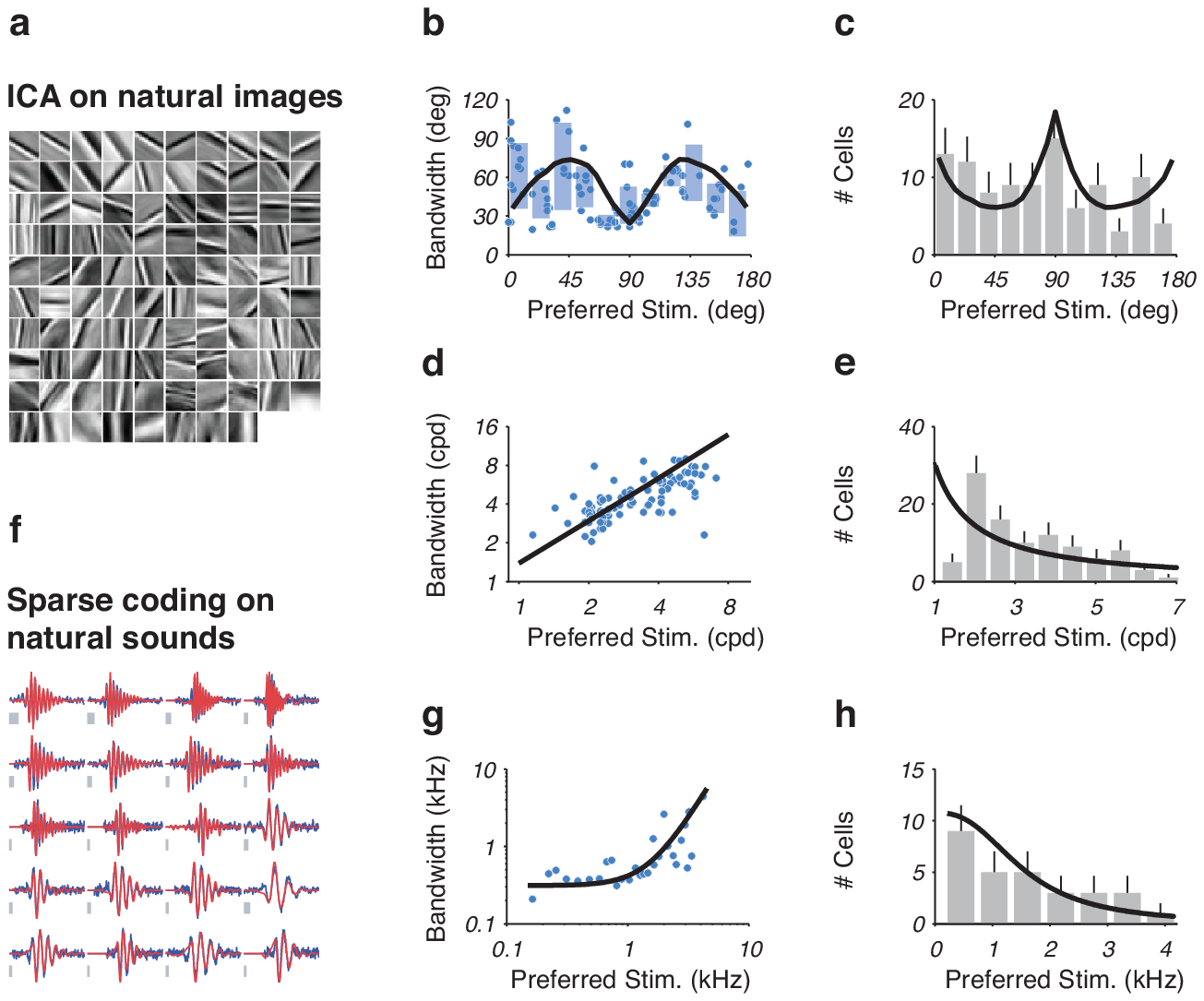}} 
\end{center}
\label{supp-effcode} \caption*{{\bf Fig A4. Comparison to ICA and sparse coding.} {\bf a}, Linear receptive fields, derived from natural images using independent components analysis (ICA)\cite{Bell1997}, resemble the receptive fields of V1 simple cells. {\bf b - e}, The tuning characteristics of this population is consistent with the predictions of our efficient coding solution (thick black lines). {\bf b}, Orientation tuning widths computed from the receptive fields in {\bf a}. Blue bars show the mean and standard deviation of the bandwidths in $16.37$ degree bins. {\bf c}, Histogram density estimate of the preferred orientations of these receptive fields. {\bf d}, Spatial frequency tuning widths computed for these receptive fields. {\bf e}, Histogram of preferred spatial frequencies of these receptive fields. Low spatial frequencies are underestimated due to the small patch sizes. {\bf f}, Linear receptive fields, derived from natural sounds (red) using sparse coding\cite{Smith2006}, resemble the receptive fields of auditory nerve fibers (blue). {\bf g-h}, Tuning characteristics of these receptive fields are consistent with the predictions of our efficient coding solution (thick black lines). {\bf g}, Acoustic frequency tuning widths\cite{Smith2006}. {\bf h}, Histogram density estimate of the preferred acoustic frequencies of the receptive fields.} 
\end{figure}

We examined the tuning properties of populations of linear receptive fields that are numerically optimized to encode ensembles of natural images or sounds, and find that they are qualitatively consistent with our physiological predictions of cell density and tuning width (Fig.~A4). Specifically, we computed populations of visual receptive fields using independent components analysis (ICA)\cite{Bell1997}, and auditory receptive fields using sparse coding \cite{Olshausen1996,Smith2006}. These optimal solutions each arise from principles that may be interpreted as variants of the efficient coding hypothesis, each using a different objective function and constraints, and making different assumptions about input distributions, neural response properties, and noise. These differences are further elaborated in the following paragraphs.

ICA is designed to recover a linear generative model of natural images,

\begin{align}
{\bf x} = {\bf A} {\bf s}, \label{genmodel} 
\end{align}

where {\bf x} is a matrix with each column representing a different natural image, {\bf A} is a matrix of basis functions, and {\bf s} is a matrix of coefficients. The ICA method aims to simultaneously estimate the basis functions, {\bf A}, and the coefficients, ${\bf s} = {\bf A}^{-1}{\bf x}$, such that the coefficients are as statistically independent as possible. Assuming no neuronal noise, this procedure is equivalent to picking the neural responses, (often equated to the coefficients ${\bf s}$), and linear receptive fields (often equated to the columns of ${\bf A}$) to maximize the information transmitted about the natural image ensemble. For our comparison, we ran the Fast ICA algorithm\cite{Hyvarinen1999} (with the default parameters) on one million $16X16$ pixel image patches sampled randomly from the same database of natural images used to compute the environmental distributions for local orientation and spatial frequency\cite{Hateren1998, Doi2003}. The dimensionality of each patch was reduced by a factor of two using principle components analysis, effectively low-pass filtering the images, in order to aid convergence. Consistent with previous literature, the optimal receptive fields \figref{A4a} are seen to closely resemble those of simple cells in V1\cite{Bell1997, Hateren1998}.

To derive the orientation and spatial frequency tuning properties of these receptive fields, we first computed the magnitude of the two dimensional Fourier transform of each filter and found the orientation and spatial frequency of the maximum amplitude. We estimated the orientation tuning curve by interpolating the values of the magnitude as a function of angle about this peak value, and the spatial frequency tuning curve by interpolating the values of the magnitude radially through the location of the peak magnitude. We found that $98$ of the $128$ receptive fields exhibited clear tuning to both orientation and spatial frequency. For these units, we computed the bandwidths of the derived tuning curves as the full width at half maximum \figref{A4b,d}. \comment{Seems unnecessary: The preferred stimuli of the tuning curves were computed as the orientation or spatial frequency that elicited the maximum response.} A histogram of the preferred stimuli was used as an estimate for the local cell density \figref{A4c,e}. For both attributes, we find that the predictions of our framework are qualitatively consistent with the derived tuning characteristics (Fig.~4b-e, thick black lines).

Sparse coding also assumes a linear generative model of the input (Eq.~\ref{genmodel}), but attempts to learn basis functions and coefficients that minimize squared reconstruction error subject to a sparsity constraint on the coefficients:

\begin{align}
\argmin_{{\bf A}, {\bf s}} \|{\bf x} - {\bf A}{\bf s} \|_{2} + \lambda \|{\bf s}\|_{0}. 
\end{align}

The first term is the mean squared error between the generative model and the inputs, and the second term is the number of non-zero coefficients (interpreted as the number of active neurons), which enforces sparsity. The parameter $\lambda$ controls the tradeoff between sparsity and reconstruction error.

When the inputs consist of natural sounds, an approximate numerical solution to this objective function yields receptive fields that resemble those of auditory nerve fibers\cite{Smith2006} \figref{A4f}. The tuning widths of these filters, as function of their preferred stimulus values, closely match the predictions of our efficient coding framework \figref{A4g}. We computed a histogram of the preferred stimuli as a local estimate of cell density and find that it is also qualitatively consistent with our results \figref{A4h}.

\subsection*{Predictions of an alternative optimality principle}
We compared our predictions to those that arise from another theory for early sensory systems $-$ that they are optimized for the {\em discrimination} of stimulus values \cite{Twer2001}. Specifically, we derived a solution for a population that minimizes the average squared stimulus discriminability. Using the Fisher bound on stimulus discriminability\cite{Series2009}, we express the average discriminability in terms of cell density and gain\cite{Ganguli2010}, yielding the following optimization problem:
\begin{equation*}
\argmax_{d(s),g(s)} \int p(s) d^{-2}(s)g^{-1}(s) \ud s, \qquad\text{subject to} \quad \int d(s) \ud s = N, \quad\text{and} \quad \int p(s)g(s) \ud s = R.
\end{equation*}

As for the case of coding efficiency, a closed form solution is readily obtained using calculus of variations:
\begin{equation}
d(s) \propto N p^{\frac{1}{2}}(s), \qquad w(s) \propto \frac{1}{d(s)} = \frac{1}{Np^{\frac{1}{2}}(s)}, \qquad g(s) \propto \frac{R}{p^{\frac{1}{2}}(s)}. \label{eq:optimum_discrimax} 
\end{equation}

The structure of this ``discrimax'' population differs significantly from that of the optimally efficient population. Specifically, the cell density is proportional to the \emph{square root} of the stimulus probability, thus allocating more cells, relative to the efficient coding solution, to less frequently occurring stimuli. The gain of the cells in the discrimax population is inversely proportional to the square root of the stimulus distribution. Since we have assumed the tuning widths are inversely proportional to cell density, and thus to the square root of the stimulus distribution, this solution again implies that the average response of each neuron (over stimuli encountered in the world), is identical across the population, as in the efficient coding solution. The unknown total resource values $\{N, R\}$ again appear only as multiplicative scale factors in the expressions for gain and density, and thus the optimal solution provides a unique and testable prediction for the shapes of both the cell density and tuning width as a function of preferred stimulus. Finally, the optimized population places limits on discrimination performance as follows:

\begin{equation}
\delta_{\rm min}(s) \propto \frac{1}{\sqrt{d^2(s) g(s)}} = \frac{1}{N \sqrt{R} p^{\frac{1}{4}}(s)}. \label{eq:discriminability_discrimax} 
\end{equation}

The solution is again a simple function of the stimulus probability, $p(s)$, scaled by a multiplicative factor that depends on neural resources and an additional factor that depends on the experimental conditions under which discrimination thresholds are measured. Although this solution predicts that frequent stimuli should be more discriminable, the form of this dependency differs significantly. \comment{Don't think we need to say this again: As a result, the solution provides a unique prediction of the shape of perceptual discriminability as a function of stimulus value.}

\begin{figure}
\begin{center}
{
\includegraphics[scale=.91]{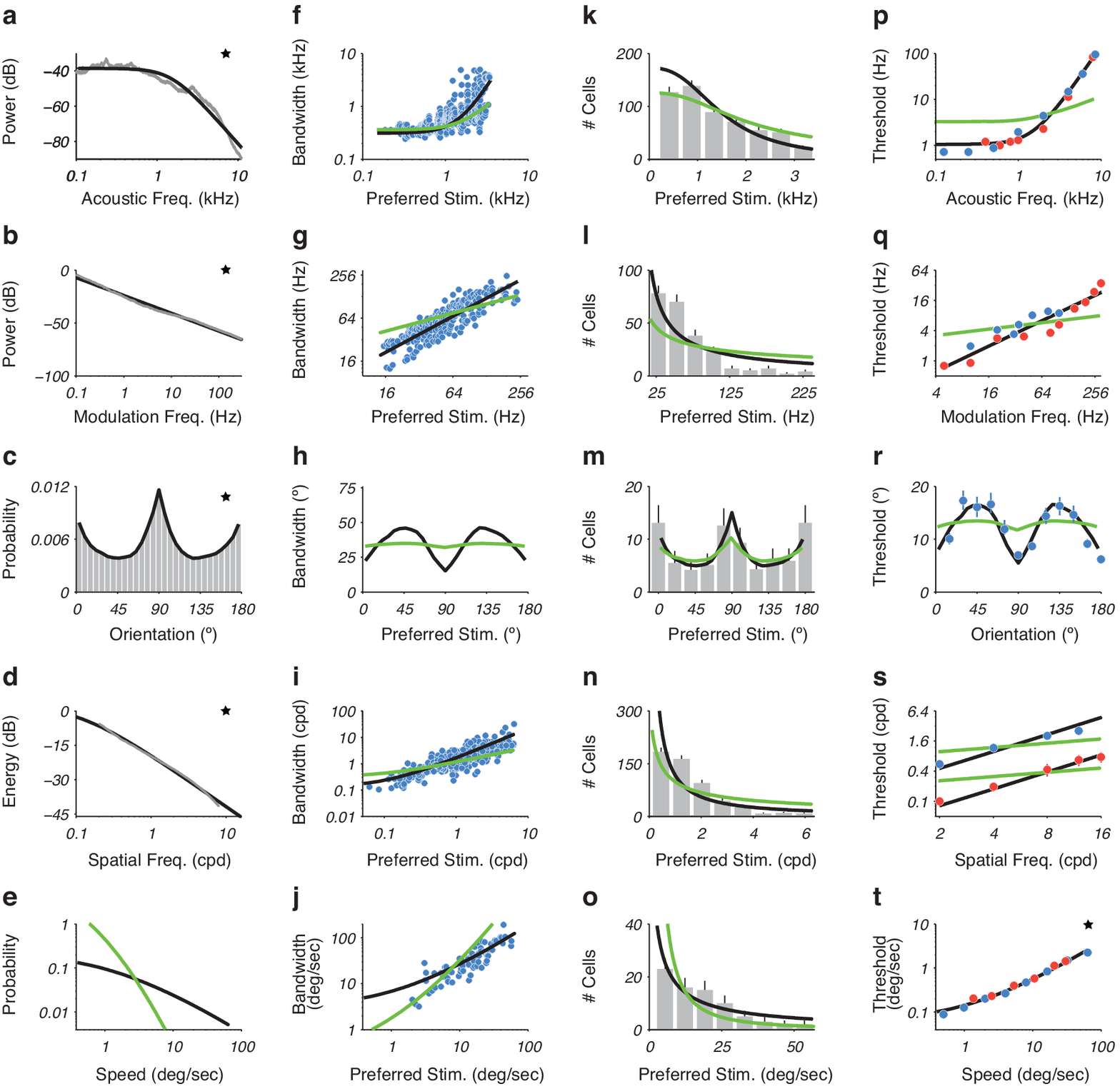}} 
\end{center}
\caption*{{\bf Fig A5. Comparison to discrimination-optimizing solution.} Predicted relationships between sensory priors, neural population properties (tuning width and cell density) and psychophysical discrimination thresholds when optimizing coding efficiency (thick black lines) and or average squared discrimination (thick green lines). Each panel is the same as in Fig.~2, except that the green curves are computed by transforming the curve in the starred panel using the discrimax solutions of Eqs. (\ref{eq:optimum_discrimax}) or (\ref{eq:discriminability_discrimax}). Each curve is rescaled to minimize the squared error of the associated data.} 
\label{supp-diffobj} 
\end{figure}

We compared these discrimax predictions to those of efficient coding \figref{A5}. For each attribute, we find that the predicted relationships between the environment, physiology, and perception for the discrimax hypothesis (Fig.~A5, thick green lines) deviate more significantly from the data than those of the efficient coding hypothesis (Fig.~A5, thick black lines). For tuning widths, the discrimax predictions account for $(32.0,55.8,36.0,55.0)\%$ of the variance in the data, which is $(7.6, 13.32, 11.7,12.4)\%$ less than that accounted for by efficient coding. The predicted cell densities also deviate significantly from data ($p<0.001$ for all attributes, one sample KS test). Finally, the new predictions of gain is inconsistent with the data, which lack any systematic relationship between the preferred stimuli and the gains (Figs.~A2 and A3c). \comment{EPS: why is it *parsimonious?? We conclude that the efficient coding hypothesis provides more parsimonious optimality principle for sensory and perceptual coding than the optimal discriminability hypothesis. }

\end{document}